# Giant anomalous Nernst signal in the antiferromagnet YbMnBi$_2$

Yu Pan[1,*], Congcong Le[1], Bin He[1], Sarah J. Watzman[1,2,3], Mengyu Yao[1], Johannes Gooth[1], Joseph P. Heremans[2,4,5], Yan Sun[1], Claudia Felser[1,*]

[1]*Max Planck Institute for Chemical Physics of Solids, Nöthnitzer Str. 40, Dresden 01187, Germany.*

[2]*Department of Mechanical and Aerospace Engineering, The Ohio State University, Columbus, OH 43210, USA.*

[3]*Department of Mechanical and Materials Engineering, University of Cincinnati, Cincinnati, OH 45221, USA.*

[4]*Department of Materials Science and Engineering, The Ohio State University, Columbus, OH 43210, USA.*

[5]*Department of Physics, The Ohio State University, Columbus, OH 43210, USA.*

*E-mail address: Yu.Pan@cpfs.mpg.de; Claudia.Felser@cpfs.mpg.de*

**Abstract**

Searching for a high anomalous Nernst effect (ANE) is crucial for thermoelectric energy conversion applications because the associated unique transverse geometry facilitates module fabrication. Topological ferromagnets with large Berry curvatures show high ANEs; however, they face drawbacks such as strong magnetic disturbances and low mobility due to high magnetization. Herein, we demonstrate that YbMnBi$_2$, a canted antiferromagnet, has a large ANE conductivity of ~10 Am$^{-1}$K$^{-1}$ that surpasses the common high values (*i.e.* 3-5 Am$^{-1}$K$^{-1}$) observed so far in ferromagnets. The canted spin structure of Mn guarantees a nonzero Berry curvature but generates only a weak magnetization three orders of magnitude lower than that of general ferromagnets. The heavy Bi with a large spin-orbit coupling enables a high ANE and low thermal conductivity, whereas its highly dispersive $p_{x/y}$ orbitals ensure low resistivity. The high anomalous transverse thermoelectric performance and extremely small magnetization makes YbMnBi$_2$ an excellent candidate for transverse thermoelectrics.

**Main**

Topological electronic structures lay the foundation for new functionalities and are crucial for various applications, including thermoelectric energy conversion.[1–3] In the past few decades, many good longitudinal thermoelectric materials have been demonstrated as topological insulators.[4,5] Recently, with the emergence of topological semimetals, the anomalous Nernst effect (ANE) has attracted increasing attention for transverse thermoelectric applications.[2,3,6–10] Owing to the large Berry curvature near the Fermi energy, high ANE thermopowers have been achieved in a few topological ferromagnets,[2,3,6–12] showing great potential for stable and precise temperature control, particularly in micro- or nanosized devices.[13,14] For example, $Co_3Sn_2S_2$,[7,8] $Co_2MnGa$[3,9], and $Fe_3Ga$[2] showed ANE thermopowers of 3-8 μV/K and ANE conductivities of 0.5-5 Am$^{-1}$K$^{-1}$, and recently $UCo_{0.8}Ru_{0.2}Al$[15] is reported to have a colossal ANE thermopower of 23 μV/K and a large ANE conductivity of 15 Am$^{-1}$K$^{-1}$. To date, high ANEs have been reported in only a few topological ferromagnets, and topological noncollinear antiferromagnets are rarely studied, except for $Mn_3X$ (X = Sn and Ge).[6,10,12] The search for large ANEs in noncollinear topological antiferromagnets will achieve advantages in broadening the material platform for transverse thermoelectrics as well as for the quest toward novel topological phenomena.

In order to utilize the ANE for practical applications, low resistivity and thermal conductivity are also required. Moreover, at the device level, low magnetization and a small inherent stray field are important for eliminating magnetic disturbances and stabilizing the remanent magnetization in the in-plane direction if using in a thin film case.[16] From these viewpoints, ANE efficiency is coupled with different parameters, similar to the longitudinal Seebeck effect,[17] and additional magnetization makes the coupling more complicated. In general, for ferromagnets, although strong ferromagnetism breaks time-reversal symmetry and offers a large Berry curvature, it also has some drawbacks. With local moment and itinerant electrons (usually due to the heavy $d$- or even $f$-bands) at the Fermi energy ($E_F$), these ferromagnets mostly have rather low mobility and high resistivity. In addition, the high magnetization and strong inherent stray field of ferromagnets introduce a strong magnetic disturbance, which cause interference with other electronic devices in practical applications.

Herein, considering the challenges outlined above, we highlight the importance of searching for materials with broken time-reversal symmetry and *p*-bands or *p-d* hybridization near the $E_F$. YbMnBi$_2$, as a canted antiferromagnet, can simultaneously realize broken time-reversal symmetry and relatively low resistivity with light bands conduction. The canted spin structure at Mn sites breaks time-reversal symmetry. Meanwhile, low resistivity can be achieved thanks to the sharp dispersion of the $p_{x/y}$ orbitals of Bi. Moreover, the large spin-orbit coupling (SOC) from Bi is critical for band topology, transverse transport response, and low thermal conductivity. Furthermore, YbMnBi$_2$ canted antiferromagnet has a much smaller magnetization and inherent stray fields than general ferromagnets. All these features demonstrate that YbMnBi$_2$ can be a new material platform beyond ferromagnets for transverse thermoelectric applications.

**Canted antiferromagnets for ANE thermoelectric applications**

A Nernst device can significantly simplify the module fabrication benefiting from its unique transverse geometry, where the device can be made simply from one material rather than needing both polarities of charge carriers in separate materials as is required of Seebeck modules. As schematically shown in **Fig. 1**(a), longitudinal Seebeck thermoelectric devices require coupled p- and n-type legs and assembled pairs.[17,18] In contrast, the Nernst device (**Fig. 1**(b)), is superior with various unique advantages that simplify the module assembly. First, complex electrical connections and associated electrical resistances are eliminated because it requires only one material. Second, in Nernst devices, the voltage can be probed in an isothermal plane, and the electrodes must be constructed only at the isothermal cold end; hence, such devices do not require contacts that are stable at high temperatures. Third, the output of a Nernst device scales extrinsically with the device size: a larger length and thickness can increase the voltage output and temperature gradient, respectively. Increasing the size in either direction would increase the associated potential drop.

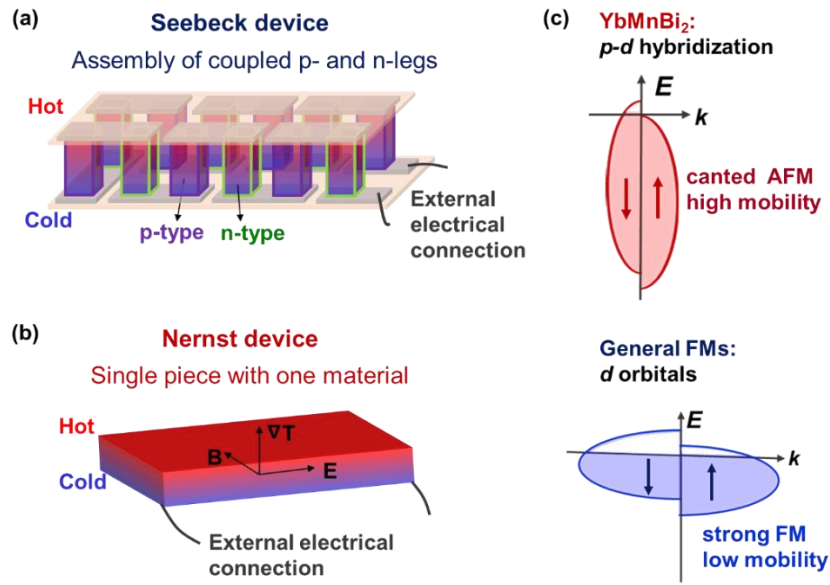

**Fig. 1** Schematic of a thermoelectric device based on the (a) Seebeck effect and (b) Nernst effect. The Nernst device needs only one material and eliminates the complex electrical connections required in the Seebeck device, especially at the hot side. (c) Schematic of the band structure near the $E_F$ for canted YbMnBi$_2$, with a comparison to general ferromagnets (FMs). YbMnBi$_2$ shows a much sharper band dispersion and a higher mobility than general FMs.

For applications as a Nernst device, YbMnBi$_2$ outshines common ferromagnets from two aspects: high mobility and low magnetization. First, as schematically shown in **Fig. 1**(c), the bands of YbMnBi$_2$ at $E_F$ are very light in the $ab$-plane, formed by the hybridization of *p* and *d* orbitals, thus guaranteeing a high mobility; whereas they are usually heavy *d* or *f* orbitals for general ferromagnets. Second, the ferromagnetism originates from the canted spin structure of Mn is extremely weak. These two advantages are superior for thermoelectric applications from the perspectives of low resistivity at the material level and small magnetic disturbance at the device level.

**The search for canted antiferromagnets**

In the search for noncollinear antiferromagnets to achieve a large ANE signal, we review the magnetic topological materials identified recently by high-throughput calculations.[19] To introduce a canted spin structure, we examine the manganese pnictides *R*Mn*Pn*$_2$, where *R* is a rare- or alkaline-earth metal, and *Pn* is a pnictide (*Pn* = P, As, Sb, or Bi). A few canted manganese pnictides—for instance, SrMnSb$_2$,[20] YbMnBi$_2$,[21] and Ca$_{1-x}$Na$_x$MnBi$_2$,[22]—have been reported to be magnetic topological materials with a canted spin structure; therefore, they are able to hold a nonzero Berry curvature.[23,24] Moreover, the bands contributed by pnictides are highly dispersive,[25] which can benefit a rather low resistivity, especially compared to that of ferromagnets.

Since Yb is relatively stable, Mn has a canted spin structure, and Bi can induce high SOC, YbMnBi$_2$ is investigated. As shown in **Fig. 2**(a), YbMnBi$_2$ crystallizes in a P4/nmm structure with two types of Bi, wherein Bi 1 is bonded with Mn, and Bi 2 forms an interlayer. The spin of Mn is antiferromagnetic along the *c*-axis but canted in the *ab*-plane. The Néel temperature was reported to be ~290 K,[26] which is very close to that observed in **Fig. 2**(b) (~283 K) and from the resistivity (**Fig. S1**). A clear "ferromagnetic" transition is observed, as shown in **Fig. 2**(b). The sharp increase in magnetization upon cooling observed in the field cooling (FC) curves indicats a spin canting temperature of ~250 K. The field dependence of magnetization is shown in **Fig. S2**. Below 250 K, the saturation magnetization value is ~$1.25 \times 10^{-3}$ $\mu_B$/f.u. in the *ab*-plane, suggesting a very small canting angle ($\theta$) of approximately 0.018°. Although the spin canting is very weak and may even be inaccessible in experiments requiring a strong signal,[26,27] it plays a crucial role in the sensitive anomalous thermal/electrical transport by inducing a nonzero Berry curvature.

Because of the canted spin structure, YbMnBi$_2$ has unique topological properties. When the SOC is ignored, a nodal line protected by glide-plane symmetry exists in the Brillouin zone owing to the band inversion of the $p_{x/y}$ orbitals of Bi 2, as shown by the yellow line in **Fig. 2**(c). When SOC is included, the nodal line is gapped, resulting in two pairs of Weyl nodes and a nonzero Berry curvature (BC). The strength of the BC depends on the $E_F$, and reaches its maximum when the $E_F$ is near the Weyl nodes. As schematically illustrated in **Fig. 2**(d), the shift in $E_F$ would probe the strength of BC, which therefore, leading to the varying strength of the AHE conductivity ($\sigma_{AHE}$) and the ANE conductivity ($\alpha_{ANE}$).[28–30] In this case, it is essential to have an optimal $E_F$ to observe a large $\sigma_{AHE}$ or $\alpha_{ANE}$, which is usually related to the quality of the single crystal.

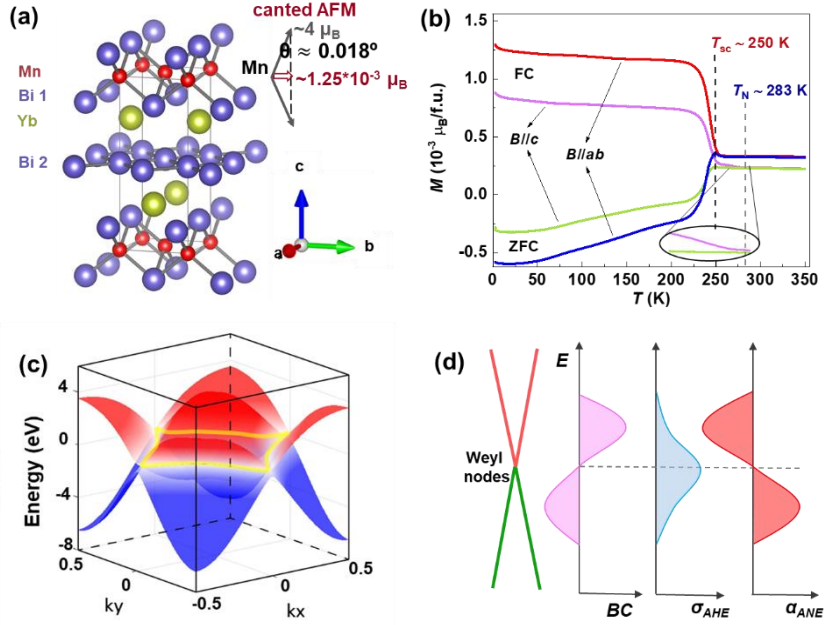

**Fig. 2** (a) Crystal structure of YbMnBi$_2$. The spin structure is canted at Mn sites, contributing to weak ferromagnetism in the *ab*-plane. (b) Temperature dependence of magnetization in the *ab*-plane and *c*-axis. Red and magenta curves denote field cooling (FC) under 1 T, whereas blue and green curves denote zero-field cooling (ZFC). The field for both FC and ZFC measurements is 100 Oe. A sharp increase in FC curves upon cooling indicates a spin canting temperature ($T_{sc}$) of ~250 K; the Néel temperature ($T_N$) is ~283 K. An extremely small canting angle of ~0.018º is resolved from the magnetization curves. (c) Nodal line (yellow line) formed near the $E_F$ in the $k_x$–$k_y$ plane. The conduction and valence bands are shown in red and blue, respectively. (d) Schematic of the $E_F$ dependence of the Berry curvature (BC), $\sigma_{AHE}$, and $\alpha_{ANE}$. The strength of the BC depends on $E_F$, resulting in the maximum $\sigma_{AHE}$ when $E_F$ is near the Weyl nodes, while the maximum $\alpha_{ANE}$ lies at different energy.

According to the theory, under a spin canting direction of (110),[21] the magnetic point group of YbMnBi$_2$ becomes *m'm*2', and only transverse transport signals in *ac*/*bc* and *ca*/*cb* can be observed owing to mirror symmetry.[31] Taking *cb* as an example, this implies that the transverse signal is measured along the *c*-axis while a temperature gradient is applied along the *b*-axis, with the magnetic field along the *a*-axis. Notably, the experimental results are highly consistent with the theoretical predictions: both ANE and AHE are observed in *bc* and *cb*, but not in *ab* (**Fig. S3** and **Fig. S4**). Therefore, the nonzero Berry curvature is believed to be a significant cause of the ANE/AHE.

**Anomalous Nernst thermopower and anomalous Hall resistivity**

The ANE thermopower ($S_{ANE}$) and AHE resistivity ($\rho_{ANE}$) (with Gerlach's sign convention, as illustrated in **Fig. S5**) in *cb* and *bc* show qualitatively identical behavior but different values. The $S_{ANE}$ and $\rho_{ANE}$ in both *cb* (**Fig. 3**(a) and (b)) and *bc* (**Fig. 3**(c) and (d)) saturates at ~1 T, which is quite small, comparable to that of $Co_2MnGa$[3,9] and much lower than $Fe_3Ga$ (~2 T).[2] The $S_{ANE}$ first increases and then decreases with heating, showing a peak at ~160 K (**Fig. 3**(e)); in contrast, the $\rho_{ANE}$ increases up to ~200 K (**Fig. 3**(f)). The maximum $S_{ANE}$ reaches a value of ~6 μV/K and ~3 μV/K at 160 K in *cb* and *bc*, respectively, which is quite high, especially for an antiferromagnet. To date, the highest ANE signals in antiferromagnets were observed in $Mn_3Sn$ (~0.5 μV/K)[6,12] and $Mn_3Ge$ (~1.2 μV/K).[10,32] The large $S_{ANE}$ are remarkable because, to the best of our knowledge, this is the first report of an antiferromagnet rivalling the ferromagnets.

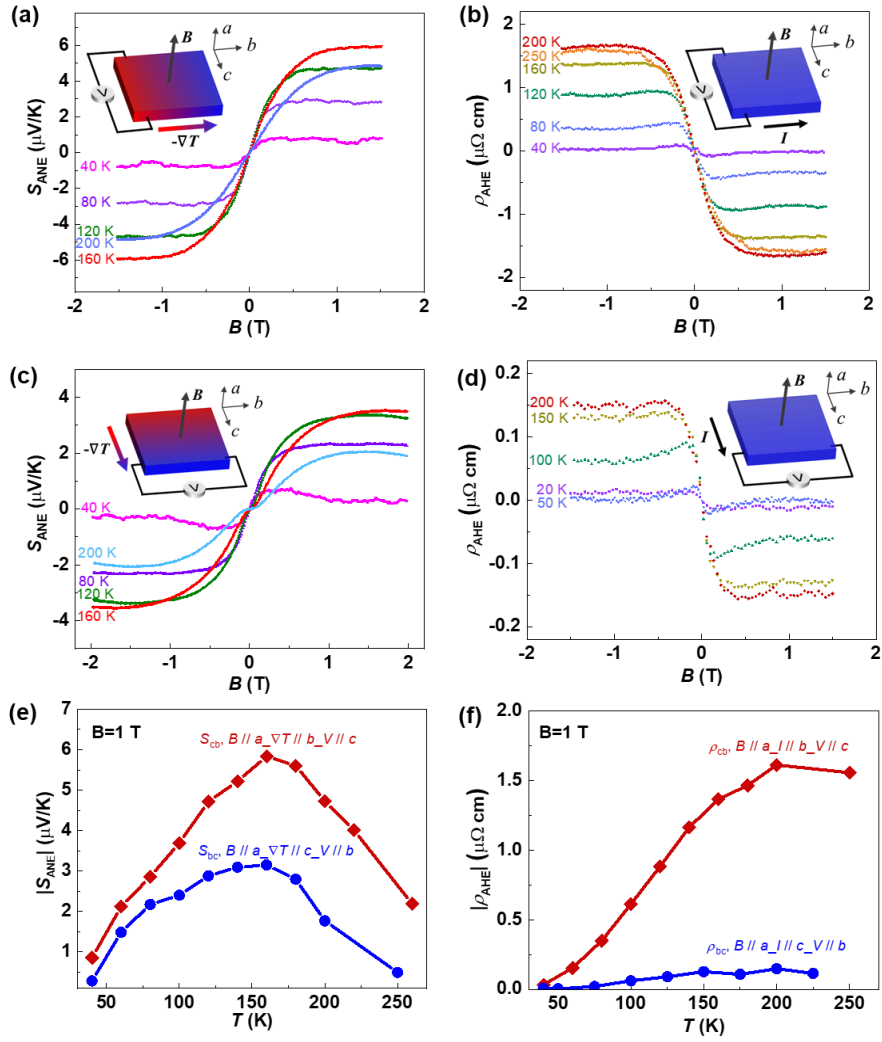

**Fig. 3** Magnetic-field dependence of ANE thermopower and AHE resistivity in (a), (b) *cb* (*V // c*) and (c), (d) *bc* (*V // b*) for various temperatures, respectively. Herein, *cb* indicates the measurement of the Nernst/Hall voltage along the *c*-axis and application of the temperature gradient along the *b*-axis, and vice versa for *bc*; the magnetic field *B* is always along the *a*-axis. The configuration of *B*, ∇*T*/*I* and *V* is schematically shown in the insets. (e), (f) Temperature dependence of the absolute values of ANE thermopower and AHE resistivity, respectively, at 1 T from 40 to 250 K.

**Intrinsic and extrinsic effects on ANE/AHE conductivity**

To quantitatively understand the origin of the ANE and AHE in YbMnBi$_2$ further, the $\alpha_{ANE}$ and $\sigma_{AHE}$ were analyzed with first-principles calculations assuming a $\theta$ of 10º. Because $\alpha_{ANE}$ is an energy derivative of $\sigma_{AHE}$,[24] we first investigated $\sigma_{AHE}$, followed by that of $\alpha_{ANE}$. Two characteristics of the band structure are essential for understanding the AHE/ANE. First, the hole pocket near the Γ point, which stems from the hybridization of the $p_z$ orbitals of Bi 1 and the $d$ orbitals of Mn, should be far below the $E_F$ if $\theta$ is zero (**Fig. S6**), but above the $E_F$ if $\theta$ is non-zero (**Fig. 4**(a)). The hole pocket near the Γ point can generate an negative Berry curvature,[31] which enhances the total Berry curvature as the electron pockets (along Γ-M, Γ-X, and Γ-Y, originating from the $p_{x/y}$ orbitals Bi 2) also generate a negative Berry curvature (**Fig. 4**(b)). As shown in **Fig. 4**(c), the Fermi surface at Γ point is observed in our crystal by angle-resolved photoemission spectroscopy (ARPES), which is highly important as it evidences the nonzero canting angle and additional negative Berry curvature from Γ band. The larger $\theta$, the more contribution generates at the Γ point, as illustrated on the right side in **Fig. 4**(c) and **Fig. S7**. **Fig. 4**(d) plots |$\sigma_{AHE}$| as a function of $\theta$; the calculated |$\sigma_{AHE}$| has a nearly linear dependence on $\theta$ from 8º to 16º. As $\sigma_{AHE}$ is zero if the canting angle is zero, we, in the simplest way, extrapolate $\sigma_{AHE}$ to zero with a linear dotted line, which therefore gives us a possibility to roughly assess the $\sigma_{AHE}$ at extremely small $\theta$.

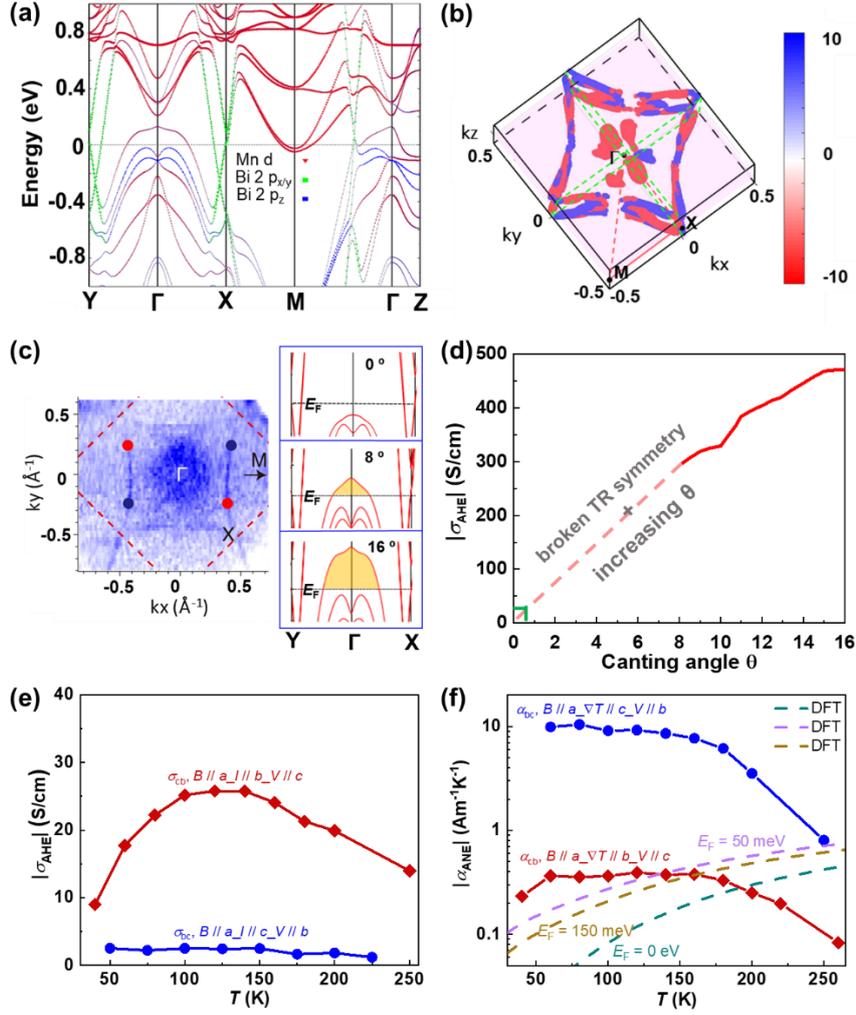

**Fig. 4** (a) Band structure of YbMnBi$_2$ and (b) Berry curvature distribution in the Brillouin zone calculated from first principles by assuming a canting angle of 10º. The total Berry curvature contribution from the electron pockets is negative. With a larger spin canting angle, the $d$ orbitals of Mn join the conduction, and an additional negative Berry curvature is generated at the Γ point. (c) Left panel: ARPES intensity at $E_F$ as a function of the $ab$-plane wave vector ($kx$ and $ky$) measured with 55 eV photons at $T$ = 19 K. The ARPES intensity is integrated over the energy range of +10 meV to −10 meV with respect to $E_F$. The inset shows two pairs of Weyl nodes. Right panel: the calculated band variation at the Γ point with increasing canting angle $θ$. (d) Calculated |$σ_{AHE}$| as a

function of $\theta$. The solid line shows the calculated results, and the dashed line shows a linear extrapolation. Experimental results of the temperature dependence of (e) $|\sigma_{AHE}|$ and (f) $|\alpha_{ANE}|$ in *cb* (*V // c*) and *bc* (*V // b*). Calculated $|\alpha_{ANE}|$ at different $E_F$ are shown by the dot lines in (f).

Second, as the electronic bands of YbMnBi$_2$ are strongly anisotropic and highly dispersive in the *ab*-plane (**Fig. S8**), there can be a large shift in $E_F$. In fact, such a shift is observed in the Hall resistivity (**Fig. S4**(c)): even an n- to p-type transition is observed from 50 to 100 K. Such high anisotropy is also observed in the Seebeck coefficient (**Fig. S9**), where positive and negative values are observed to coexist in different crystallographic directions, which may indicate that YbMnBi$_2$ is a goniopolar material.[33] This makes it challenging to deal with the actual $E_F$ in specific directions as it can even be different in different directions. Based on the analysis above, we can attain a better understanding of $|\sigma_{AHE}|$ and $|\alpha_{ANE}|$.

As shown in **Fig. 4**(e), the maximum experimental $|\sigma_{AHE}|$ may be coordinated to the theoretically predicted result with a very small $\theta$ (as illustrated by the point indicated with green lines in **Fig. 4**(d)). The decreasing trend above ~150 K can be attributed to the temperature disturbance on the spin structure; however, the increasing trend below 150 K is rather uncommon. $|\alpha_{ANE}|$ also shows an unusual decreasing trend, as shown in **Fig. 4**(f). In addition, anisotropic behaviors are observed in $\sigma_{AHE}$ and $\alpha_{ANE}$. Akgoz and Saunders[34,35] combined the reciprocity relations with the crystal symmetries to derive the allowed symmetry operations in various point groups. The magnetic point group of YbMnBi$_2$ being *m'm2'*, one expects that the Hall resistivity $\rho_{bc}(H_a) = -\rho_{cb}(-H_a)$,[34] and the same relation would be expected to hold for the conductivity tensors. The fact that this is not observed in **Fig. 4**(e) points to the absence of time-reversal symmetry and the anomalous origin of the Hall conductivities. For the thermoelectric conductivity tensor, the observation that $S_{bc}(H_a) \neq -S_{cb}(-H_a)$ (**Fig. 3**(e)) and $\alpha_{bc}(H_a) \neq -\alpha_{cb}(-H_a)$ (**Fig. 4**(f)) is allowed.[35] Most importantly, a maximum $|\alpha_{ANE}|$ of 10 Am$^{-1}$K$^{-1}$ is achieved in *bc*, which is much higher than those of most of the ferromagnets, including Co$_3$Sn$_2$S$_2$,[7] Co$_2$MnGa,[9] Fe$_3$Ga,[2] and SmCo$_5$,[36] whose record value is ~5 Am$^{-1}$K$^{-1}$, and even one order higher than those of the antiferromagnets Mn$_3$X (X = Sn[6,12] and Ge [10,32]).

There can be extrinsic contributions to such a large $\alpha_{ANE}$ in addition to the Berry curvature, particularly considering three aspects. First, there is a mismatch between the experimental and the first-principles predicted results. The calculated $|\alpha_{ANE}|$ is much lower than the experimental $\alpha_{bc}$, even by tuning the $E_F$ (**Fig. S10**). The negative temperature dependence of $|\alpha_{ANE}|$ is similar to that recently been observed in MnBi, which is attributed to an extrinsic magnon-drag effect and can account for the deviation in temperature dependence between experiment and theory.[37] Second, anisotropic behavior is observed in $\alpha_{ANE}$, *i.e.* $\alpha_{bc} \neq -\alpha_{cb}$. In principle, from DFT, $\alpha_{bc}^{intrinsic} = -\alpha_{cb}^{intrinsic}$ only if it is solely originated from the Berry curvature, as determined by the Berry phase of the wavefunction, $\Omega_{ij}(\vec{k}) = -\Omega_{ji}(\vec{k})$. Third, a large violation of the ratio of $|\alpha_{bc}/\sigma_{bc}|$ from an empirical value of $k_B/e$ is observed (**Fig. S11**). As recently argued by Behnia *et al.*, the ratio of $|\alpha_{ANE}/\sigma_{ANE}|$ saturates to ~$k_B/e$ at 300 K if both are originated from intrinsic Berry curvature.[8,38]

In fact, extrinsic contributions are possible in real crystals due to the unavoidable presence of scattering, particularly for large SOC systems, which is the case for YbMnBi$_2$. The giant SOC induces an additional skew force on the spin polarized charge carriers.[39,40] Previous materials with large ANEs, including Co$_3$Sn$_2$S$_2$,[7] Co$_2$MnGa,[9] Fe$_3$Ga,[2] and Mn$_3$X (X = Sn and Ge),[6,10,12,32] have a small SOC without heavy elements, which is distinctly not the case here. Very recently, Fu *et al.* indicated that extrinsic contributions can be significant, and even potentially dominant over Berry curvature. For accurately describing $\alpha_{ANE}$, significant improvements to current modeling techniques, beyond the simple single band models, are necessary. These improvements include, but are not limited to, band anisotropy, deviations from a linear band dispersion, and interactions of point defects with electrons.[41] Notably, the strong band anisotropy in YbMnBi$_2$ have been rarely observed in other materials with large ANE. We highlight that the significant anisotropy in the Fermi surface of YbMnBi$_2$ plays a significant role in achieving large $\alpha_{ANE}$ in *bc*, while not in *cb*. The electron pockets are highly dispersive in *ab*-plane but have nearly no dispersion along *c*-axis, leading to a great difference in the extrinsic scattering effects on charge carriers between *bc* and *cb*. The extremely high $\sigma_{bb}$ (which is ~35 times higher than $\sigma_{cc}$ at 80 K), along with asymmetric skew scattering rates, results in high $\alpha_{ANE}$ only in *bc*. Future deeper investigations are sufficiently warranted to explore both the intrinsic (*e.g.* developing a detailed distribution of the Berry curvature in the Brillouin zone) and extrinsic contributions (*e.g.* strength of skew scattering and side jump related to SOC) to the large $\alpha_{ANE}$.

**Beyond the ANE performance**

In addition to the large ANE signals, YbMnBi$_2$ presents extremely small $M$, which can reduce the magnetic field interaction on surrounding electronic devices during practical applications. **Fig. 5**(a) compares $|\alpha_{ANE}|/M$, $|S_{ANE}|/M$, and $M$ of YbMnBi$_2$ to those of other compounds with high $S_{ANE}$—namely, the ferromagnets Fe$_3$Ga,[2] Co$_2$MnGa,[3,9] Co$_3$Sn$_2$S$_2$,[7] UCo$_{0.8}$Ru$_{0.2}$Al,[15] MnBi,[37], Ga$_{1-x}$Mn$_x$As,[42] and the chiral antiferromagnets Mn$_3$Sn[6,12] and Mn$_3$Ge.[10,32] Clearly, the noncollinear antiferromagnets, especially YbMnBi$_2$, surpass all ferromagnets in terms of $|\alpha_{ANE}|/M$ and $|S_{ANE}|/M$, demonstrating that canted antiferromagnets, even with extremely low magnetization, can show a large ANE that competes with the best ferromagnets. For practical applications, using permanent magnets (with which 0.3 T is common, 1 T is possible by carefully choosing materials) is reasonable for materials with no remnant magnetization, such as YbMnBi$_2$. Further exploring the zero-field ANEs of polycrystals and/or thin films would be of interest for future devices.

In addition to the large ANE signals, because of the anisotropy, the resistivity and thermal conductivity can be decoupled to realize simultaneous gains in both—*i.e.*, low $\rho_{bb}$ and low $\kappa_{cc}$ (**Fig. S1**). Both $\rho_{bb}$ and $\kappa_{cc}$ are much lower than those of the ferromagnets with a large ANE (**Table S1**). Consequently, YbMnBi$_2$ shows high anomalous Nernst thermoelectric figure of merit $zT_{ANE}$, defined as, $zT_{ANE,bc} = S_{ANE,bc}^2 T/(\rho_{bb}\kappa_{cc})$. As shown in **Fig. 5**(b), YbMnBi$_2$ has the highest $zT_{ANE}$ in a wide temperature range, demonstrating that YbMnBi$_2$ has a better comprehensive anomalous transverse thermoelectric performance than any other materials so far. Though the $zT_{ANE}$ values are still significantly lower than the traditional, longitudinal $zT$, we present this result as promising progress for using ANE in thermoelectric applications, which could ultimately offer improvements for microscale devices and the field of spin caloritronics.

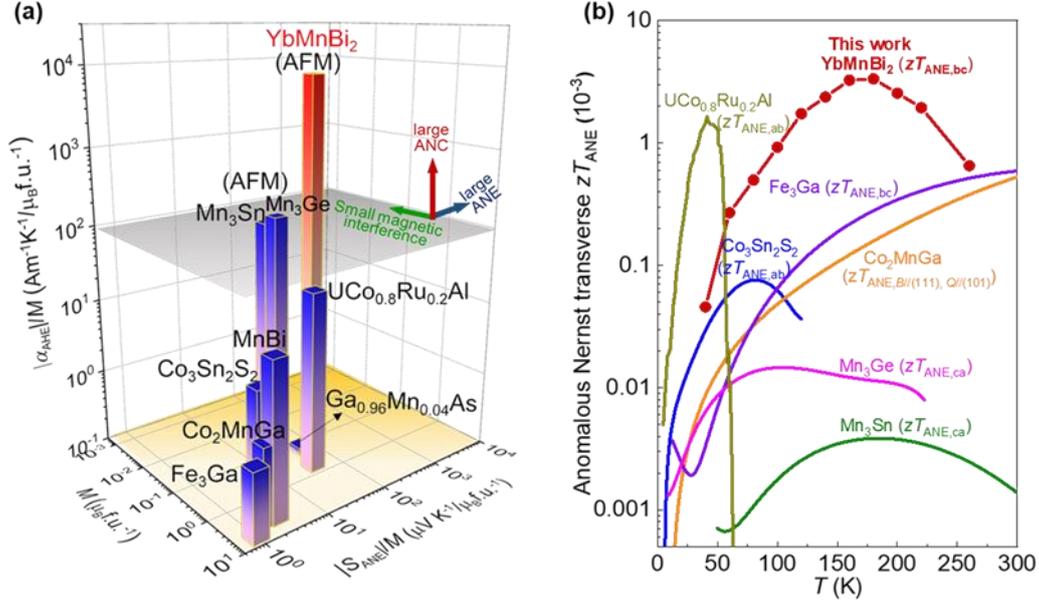

**Fig. 5** (a) Comparison of the absolute values of $|\alpha_{ANE}|/M$, $|S_{ANE}|/M$, and $M$ of YbMnBi$_2$ in $bc$ (160 K) to those of other compounds with high ANE thermopowers—namely, the ferromagnets Fe$_3$Ga (300 K),[2] Co$_2$MnGa (300 K),[3,9] and Co$_3$Sn$_2$S$_2$ (80 K),[7] UCo$_{0.8}$Ru$_{0.2}$Al (40 K),[15] MnBi (80 K),[37] Ga$_{1-x}$Mn$_x$As (10 K),[42] and the chiral antiferromagnets Mn$_3$Sn (300 K)[6,12] and Mn$_3$Ge (300 K).[10,32] $M$ is the magnetization of the compound. (b) Temperature dependence of the anomalous Nernst thermoelectric figure of merit $zT_{ANE}$ values of YbMnBi$_2$ in $bc$ configuration, with a comparison to those of other compounds with high ANE thermopowers.

**Summary and outlook**

In summary, YbMnBi$_2$, a canted antiferromagnet, breaks time-reversal symmetry and yields a nonzero Berry curvature. Together with a large SOC contributed by the heavy Bi atoms, a high ANE conductivity of 10 Am$^{-1}$K$^{-1}$ is observed, which is even 2-3 times higher than that of most of the ferromagnets and is the best reported so far among noncollinear antiferromagnets, which is of great interest, particularly considering its small magnetization. Furthermore, it shows strong anisotropy, making it possible to achieve low resistivity and low thermal conductivity simultaneously in transverse thermoelectric configurations. All these unique features demonstrate the great potential of YbMnBi$_2$ as a transverse thermoelectric material.

The present results suggest that canted antiferromagnets, including but not limited to the *R*Mn*Pn*$_2$ family, are worthy of further investigation for large ANEs. Future studies searching for compounds with canting temperatures higher than room temperature are crucial. Moreover, investigating the materials in polycrystalline or thin film cases to induce remanent magnetizations can be important for practical applications without external magnetic field. In addition, a precise tuning of the Fermi level would be efficient to enhance the ANE. Finally, with a deeper understanding of the ANE from both intrinsic and extrinsic contributions experimentally and theoretically in the future, we expect boosted ANE in more topological materials.

**Methods**

*Ab-initio calculations*: Calculations were conducted using density functional theory (DFT) implemented in the Vienna *ab initio* simulation package (VASP) code.[43,44] To calculate the band structure, the Perdew–Burke–Ernzerhof (PBE) exchange-correlation functional and projector-augmented-wave approach were used. The cut-off energy was set as 500 eV to expand the wave functions into a plane-wave basis. The Brillouin zone was sampled in the *k* space within the Monkhorst–Pack scheme,[45] and a *k* mesh of 10

× 10 × 4 was adopted based on an equilibrium structure. The C-type antiferromagnetic order along the *c*-axis at the Mn sites, a canted spin structure, and SOC were considered. To calculate the anomalous Hall conductivity and anomalous Nernst conductivity, the *ab initio* DFT Bloch wave function was projected onto highly symmetric atomic-orbital-like Wannier functions[46] with a diagonal position operator using VASP code.[43,44] To obtain precise Wannier functions, we included the outermost *s*- and *d*-orbital for Yb, *d*-orbital for Mn, and *p*-orbital for Bi to cover the full band overlap from the *ab initio* and Wannier functions.

*Sample preparation*: YbMnBi$_2$ single crystals were grown using a self-flux method with an elemental ratio of Yb:Mn:Bi=1:1:4.[25] Yb (99.99%), Mn (99.98%), and Bi (99.999%) were cut into small pieces and mixed before being placed in an alumina crucible. Subsequently, they were sealed in a quartz tube under a partial argon pressure. The sealed tube was then heated to 1050 °C in 2.5 days and maintained for 24 h. Next, it was slowly cooled to 400 °C at a rate of 2 °C/h, and single crystals were subsequently obtained by removing the flux through centrifugation.

*Sample characterization*: The single crystallinity and orientation of the as-grown single crystal were determined using the Laue X-ray diffraction (**Fig. S12**). Composition and homogeneity were examined by scanning electron microscopy (Philips XL30) with an Oxford energy-dispersive X-ray spectroscopy (EDX, Quantax, Bruker) (**Fig. S13**).

*ARPES measurements*: The ARPES experiment was performed at the Bloch beamline at MAX IV with a Scienta DA30 analyzer. The sample was cleaved in situ at 19 K with a base pressure lower than $1 \times 10^{-10}$ mbar.

*Measurement of transport properties*: Resistivities and Hall resistivities were measured using a Physical Property Measurement System (PPMS9, Quantum Design) in an Electrical Transport Option via a standard four-probe method. The Nernst thermopower, Seebeck coefficient, and thermal conductivity were measured in the PPMS under a high vacuum by using a standard four-contact steady-state method.[47] The magnetization was measured using a Magnetic Property Measurement System (MPMS3, Quantum Design).

**Acknowledgements**

This work was supported by the Deutsche Forschungsgemeinschaft (DFG, German Research Foundation)-Projektnummer (392228380), the ERC Advanced Grant No. (742068) "TOP-MAT", and the European Union's Horizon 2020 research and innovation programme (no. 766566) "ASPIN". YP acknowledges financial support from the Alexander von Humboldt Foundation. SJW acknowledges support from the U.S. Department of Energy, Office of Science, Office of Basic Energy Sciences Early Career Research Program Award Number DE-SC0020154. JPH acknowledges support from the U.S. Army Research Office grant W911NF2120089. We acknowledge Balasubramanian Thiagarajan and Craig Polley for their help in the ARPES experiments.

**Author contributions**

Y.P. grew the single crystal and conducted the crystallinity and composition characterization. Y.P. and S.J.W. designed the experiments. Y.P. and B.H. measured the transport properties. C.L. and Y.S. completed the theoretical calculations. M.Y.

performed the ARPES experiments. J.P.H. and C.F. supervised the project. All authors discussed the results and contributed to the manuscript preparation.

**Figure Legends**

**Fig. 1** Schematic of a thermoelectric device based on the (a) Seebeck effect and (b) Nernst effect. The Nernst device needs only one material and eliminates the complex electrical connections required in the Seebeck device, especially at the hot side. (c) Schematic of the band structure near the $E_F$ for canted YbMnBi$_2$, with a comparison to general ferromagnets (FMs). YbMnBi$_2$ shows a much sharper band dispersion and a higher mobility than general FMs.

**Fig. 2** (a) Crystal structure of YbMnBi$_2$. The spin structure is canted at Mn sites, contributing to weak ferromagnetism in the *ab*-plane. (b) Temperature dependence of magnetization in the *ab*-plane and *c*-axis. Red and magenta curves denote field cooling (FC) under 1 T, whereas blue and green curves denote zero-field cooling (ZFC). The field for both FC and ZFC measurements is 100 Oe. A sharp increase in FC curves upon cooling indicates a spin canting temperature ($T_{sc}$) of ~250 K; the Néel temperature ($T_N$) is ~283 K. An extremely small canting angle of ~0.018º is resolved from the magnetization curves. (c) Nodal line (yellow line) formed near the $E_F$ in the $k_x$–$k_y$ plane. The conduction and valence bands are shown in red and blue, respectively. (d) Schematic of the $E_F$ dependence of the Berry curvature (BC), $\sigma_{AHE}$, and $\alpha_{ANE}$. The strength of the BC depends on $E_F$, resulting in the maximum $\sigma_{AHE}$ when $E_F$ is near the Weyl nodes, while the maximum $\alpha_{ANE}$ lies at different energy.

**Fig. 3** Magnetic-field dependence of ANE thermopower and AHE resistivity in (a), (b)

*cb* (*V* // *c*) and (c), (d) *bc* (*V* // *b*) for various temperatures, respectively. Herein, *cb* indicates the measurement of the Nernst/Hall voltage along the *c*-axis and application of the temperature gradient along the *b*-axis, and vice versa for *bc*; the magnetic field *B* is always along the *a*-axis. The configuration of *B*, $\nabla T/I$ and *V* is schematically shown in the insets. (e), (f) Temperature dependence of the absolute values of ANE thermopower and AHE resistivity, respectively, at 1 T from 40 to 250 K.

**Fig. 4** (a) Band structure of YbMnBi$_2$ and (b) Berry curvature distribution in the Brillouin zone calculated from first principles by assuming a canting angle of 10º. The total Berry curvature contribution from the electron pockets is negative. With a larger spin canting angle, the *d* orbitals of Mn join the conduction, and an additional negative Berry curvature is generated at the Γ point. (c) Left panel: ARPES intensity at $E_F$ as a function of the *ab*-plane wave vector (*kx* and *ky*) measured with 55 eV photons at *T* = 19 K. The ARPES intensity is integrated over the energy range of +10 meV to −10 meV with respect to $E_F$. The inset shows two pairs of Weyl nodes. Right panel: the calculated band variation at the Γ point with increasing canting angle *θ*. (d) Calculated |$\sigma_{AHE}$| as a function of *θ*. The solid line shows the calculated results, and the dashed line shows a linear extrapolation. Experimental results of the temperature dependence of (e) |$\sigma_{AHE}$| and (f) |$\alpha_{ANE}$| in *cb* (*V* // *c*) and *bc* (*V* // *b*). Calculated |$\alpha_{ANE}$| at different $E_F$ are shown by the dot lines in (f).

**Fig. 5** (a) Comparison of the absolute values of |$\alpha_{ANE}$|/*M*, |$S_{ANE}$|/*M*, and *M* of YbMnBi$_2$ in *bc* (160 K) to those of other compounds with high ANE thermopowers—namely, the ferromagnets Fe$_3$Ga (300 K),[2] Co$_2$MnGa (300 K),[3,9] and Co$_3$Sn$_2$S$_2$ (80 K), [7]

UCo$_{0.8}$Ru$_{0.2}$Al (40 K),[15] MnBi (80 K),[37] Ga$_{1-x}$Mn$_x$As (10 K),[42] and the chiral antiferromagnets Mn$_3$Sn (300 K) [6,12] and Mn$_3$Ge (300 K).[10,32] *M* is the magnetization of the compound. (b) Temperature dependence of the anomalous Nernst thermoelectric figure of merit *zT*$_{ANE}$ values of YbMnBi$_2$ in *bc* configuration, with a comparison to those of other compounds with high ANE thermopowers.